# Spin Gain Transistor in Ferromagnetic Semiconductors: the Semiconductor Bloch Equations Approach

Dmitri E. Nikonov and George I. Bourianoff

Intel Corp., SC1-05, 3065 Bowers Ave., Santa Clara, CA 95054

Abstract

Scheme and principle of operation of a "spin gain transistor" are proposed: a large unmagnetized current creates density sufficient for the ferromagnetic transition; a small magnetized current initiates spontaneous magnetization; large magnetized current is extracted. Therefore spin gain of >1000 is predicted. Collective dynamics of spins under Coulomb exchange interaction is described via Semiconductor Bloch Equations.

# I Introduction

Recent scientific breakthroughs, such as efficient injection of spins from a ferromagnet to a semiconductor [1], long-lived spin coherence in a semiconductor, electric and optical control of such spins [2,3,4], high-temperature ferromagnetism in magnetically-doped III-V semiconductors [5], and carrier control of ferromagnetism [6,7], led to a vigorous growth of the emergent field of semiconductor spintronics.

Various semiconductor logic devices that are using the spin degree of freedom for their operation have been proposed: the original spin modulator by Datta and Das [8], spin filter [9], unipolar spin transistor [10], semiconductor magnetic bipolar transistor [11,12] and many others. The commonality of these devices is that they utilize some form of dependence of carrier transport (conduction, injection, relaxation, etc.) on spin in order to control the current through the device. The spin state in its turn is controlled by the magnetization of a ferromagnetic contact or by the external electromagnetic field. In this approach one can at best increase the spin polarization degree of an input current or to filter a fraction of the current based on its spin.

We believe that the future of spintronics depends on its ability to demonstrate a device with spin gain, akin to an electronic transistor with gain where a gate (or base) current controls a larger source-drain (or emitter-collector) current. Similarly, a small current with maybe only a small spin polarization must create a large spin polarization in a large output current; it also must control the direction of the polarization. (The definitions and the reasoning are quantified below.)

In the device we propose, which we call "spin gain transistor", spin gain is achieved via creation of conditions for the ferromagnetic transition by injecting enough carriers, and then switching a small spin-polarized control current which acts to break the isotropy and to induce spontaneous magnetization in the same direction as the control one. Note that no external magnetic field is required for operation of the device. Voltage control of magnetism in a field-effect transistor has been discussed in Ref. [13]. Our device shares some similarities with magnetic bipolar transistors [11][12]. The difference is that our device operates with just majority carriers; the base is not initially magnetized, its magnetization is controlled by a small base current. Also our scheme does not require an optical field to control ferromagnetism in the base. It is also different from the unipolar spin transistor [10] since it does not use junctions between regions with opposite magnetizations.

The article is organized as follows. In Section II we qualitatively describe the underlying physical phenomena and the principle of operation of the proposed device. In section III a detailed description of the mathematical model for the device is provided. We develop a theory of the spin dynamics in the transistor based on the Semiconductor Bloch equations [14] containing Coulomb interaction between carriers. We show that the equations are capable of describing the collective ferromagnetic state of the spins. Various relaxation processes, injection and removal of carriers by current are naturally incorporated in this approach, while it would be difficult in other many-body techniques, such as the mean-filed theory [15], Green's function approach [16], or density functional approach [17]. Though high ferromagnetic transition temperatures ("Curie temperatures") have been achieved in semiconductors via the interaction of free carriers with localized spins of magnetic impurities, such as Mn, in this work we deliberately keep the model of ferromagnetism simple, including just the carrier interactions. The theory of realistic magnetic semiconductors would, of course, require lifting this limitation. Section IV contains the numerical results and their discussion: evolution of spins in the magnetic field, state of the spontaneous ferromagnetism without the field, time-dependent description of the transistor operation. The article is summarized in Section V.

## II. Principle of operation

The operation of the spin transistor is based on the following effects of carrier control of ferromagnetism, which will be quantitatively described later. In a semiconductor, at a sufficiently high temperature and sufficiently low density, the spins of all free carriers are randomly polarized. Free carriers interact via the electrostatic Coulomb force. Due to the fermionic nature of free carriers, the energy of Coulomb interaction is the lowest when all spins are aligned in the same direction. Carriers do in fact form such a ferromagnetic state at a sufficiently low temperature. This state can exist spontaneously, i.e., without an external magnetic field.

The band diagram of the structure is presented in Figure 1a, band bending is not shown for simplicity. The transistor consists of a p-doped emitter and collector and an undoped base. The same principle is applicable to reversed polarities of doping and currents. Here for simplicity we discuss operation with only one polarity of carriers. One can also envision operation involving both electrons and holes. In this case the structure of the transistor is reminiscent of a magnetic bipolar transistor [11].

In the initial state, Figure 1a, none of the regions are magnetized. The density of holes in the base region is below the ferromagnetic transition for the given temperature. Therefore a small spin-polarized base current cannot produce magnetization in the base. The cycle starts with applying the voltage to the emitter to inject the current, Figure 1b. After this stage the density in the base is sufficient for a ferromagnetic state to form, but since no factor breaks the isotropy, the spins are still randomly polarized. Next a small spin-polarized current is injected to the base, Figure 1c. The spin polarized current can originate from a ferromagnetic metal or it can be an output from another spin transistor in the integrated circuit. The current breaks the isotropy and causes the spins of all holes to align with its spin direction. Finally, Figure 1d, the voltage is applied to the collector to cause a current from the base which is strongly spin polarized. The nature of the spin gain transistor is the control of this output spin current by a small input current. The ratio of the polarized components of the two current will be termed "spin gain".

To avoid unnecessary complications, we will model only the distribution of carriers in the base and consider it spatially-uniform. Electric currents will be described as the exchange of particles with reservoirs in a thermal equilibrium at ambient temperature. The reservoirs correspond to the emitter, collector, and the base contact. Further simulations will incorporate realistic spatial dependence of carrier distributions in the transistor. The base in this article will be considered a quantum well, i.e., the carriers will be confined in two spatial dimensions.

## III. Mathematical model

In this section we describe the equations used to model the dynamics of spins of free charge carriers in a spin transistor under the action of a magnetic field and with Coulomb interaction between them and with incoming or outgoing spin-polarized current.

**Spin algebra**

In this subsection we restate for reference the formulas describing the spin states, see e.g. [18]. Electrons have the spin of $s = 1/2$ (in units of the Plank's constant $\hbar$) and therefore its spin state is described in a basis of $2s + 1 = 2$ states. The basis can be chosen as eigenstates of a projection of spin on any axis, e.g. z-axis, with eigenvalues $s_z = 1/2$ and $s_z = -1/2$, which we will call "spin-up" $|u\rangle$ and "spin-down" $|d\rangle$ states. These states in the matrix notation are written as

$$|u\rangle = \begin{pmatrix} 1 \\ 0 \end{pmatrix}, \qquad |d\rangle = \begin{pmatrix} 0 \\ 1 \end{pmatrix}. \tag{1}$$

Note that the isotropic nature of the description is still preserved despite the choice of the axis to project the spin, until some external conditions, such as magnetic field, break the symmetry. The wavefunction of any pure state of the spin can be expressed via "ket" vectors or via matrix notation

$$|\psi\rangle = \alpha |u\rangle + \beta |d\rangle, \qquad \psi = \begin{pmatrix} \alpha \\ \beta \end{pmatrix}. \tag{2}$$

Then the complex conjugate wavefunction is expressed via "bra" vectors or via Hermitean-conjugate matrices

$$\langle\psi| = \alpha^* \langle u| + \beta^* \langle d|, \qquad \psi^* = \begin{pmatrix} \alpha^* & \beta^* \end{pmatrix}. \tag{3}$$

The scalar product of two vectors expresses the overlap of the two wavefunctions

$$(\psi_1, \psi_2) \equiv \langle\psi_1|\psi_2\rangle = \begin{pmatrix} \alpha_1^* & \beta_1^* \end{pmatrix} \begin{pmatrix} \alpha_2 \\ \beta_2 \end{pmatrix} = \alpha_1^* \alpha_2 + \beta_1^* \beta_2. \tag{4}$$

Note that the scalar product can also be alternatively written with the interchanged places of the wavefunctions

$$(\psi_1, \psi_2) \equiv \mathrm{Tr}\big[|\psi_2\rangle\langle\psi_1|\big] = \mathrm{Tr}\left[\begin{pmatrix}\alpha_1^* \\ \beta_1^*\end{pmatrix}(\alpha_2 \quad \beta_2)\right]. \tag{5}$$

The condition of unity probability in a complete basis of states dictates the normalization

$$\langle\psi|\psi\rangle = 1. \tag{6}$$

The basis states are obviously both normalized and orthogonal

$$\langle u|d\rangle = 0. \tag{7}$$

Any operator $\hat{O}$ (which corresponds to a physical quantity) is represented by a matrix

$$\hat{O} = \begin{pmatrix} O_{uu} & O_{ud} \\ O_{du} & O_{dd} \end{pmatrix}. \tag{8}$$

Therefore the operator can be represented via the basis of "bra" and "ket" states as follows

$$\hat{O} = O_{uu}|u\rangle\langle u| + O_{ud}|u\rangle\langle d| + O_{du}|d\rangle\langle u| + O_{dd}|d\rangle\langle d|. \tag{9}$$

Eigenvalues $O_n$ and eigenfunctions $|\psi_n\rangle$ of operators are defined via the equation

$$\hat{O}|\psi_n\rangle \equiv \begin{pmatrix} O_{uu} & O_{ud} \\ O_{du} & O_{dd} \end{pmatrix}\begin{pmatrix}\alpha_n \\ \beta_n\end{pmatrix} = O_n\begin{pmatrix}\alpha_n \\ \beta_n\end{pmatrix} \equiv O_n|\psi_n\rangle. \tag{10}$$

Examples of operators in the spin space are the real-space projections of a spin vector

$$\mathbf{S} = \frac{\hbar}{2}\boldsymbol{\sigma}. \tag{11}$$

They are given by the Pauli matrices corresponding to each axis

$$\sigma_x = \begin{pmatrix} 0 & 1 \\ 1 & 0 \end{pmatrix}, \qquad \sigma_y = \begin{pmatrix} 0 & -i \\ i & 0 \end{pmatrix}, \qquad \sigma_z = \begin{pmatrix} 1 & 0 \\ 0 & -1 \end{pmatrix}. \tag{12}$$

Easy to see that eigenvectors of the Pauli matrices, i.e., quantum states with a defined value of spin projection on an axis, can be chosen as in the following examples

$$|z,+1\rangle = \begin{pmatrix} 1 \\ 0 \end{pmatrix}, \quad |x,-1\rangle = \frac{1}{\sqrt{2}}\begin{pmatrix} 1 \\ -1 \end{pmatrix}, \quad |y,+1\rangle = \frac{1}{\sqrt{2}}\begin{pmatrix} 1 \\ i \end{pmatrix}. \tag{13}$$

The average value of the physical quantity associated with the operator is calculated, in a pure state, as

$$\langle O \rangle = (\psi, \hat{O}\psi) \equiv \langle \psi | \hat{O} | \psi \rangle = (\alpha^* \quad \beta^*) \begin{pmatrix} O_{uu} & O_{ud} \\ O_{du} & O_{dd} \end{pmatrix} \begin{pmatrix} \alpha \\ \beta \end{pmatrix}. \tag{14}$$

Alternatively, using the property (5), it can be written via the density matrix $\rho$ [18]:

$$\langle O \rangle = Tr\left[\hat{O}|\psi\rangle\langle\psi|\right] \equiv Tr\left[\hat{O}\rho\right]. \tag{15}$$

Mixed states are weighted averages of many pure states with probabilities $P(\psi)$ and thus their density matrix [19] is

$$\rho = \sum_{\psi} P(\psi) |\psi\rangle\langle\psi|. \tag{16}$$

In the matrix notation and in the general mixed state the density matrix is

$$\rho = \begin{pmatrix} \rho_{uu} & \rho_{ud} \\ \rho_{du} & \rho_{dd} \end{pmatrix}. \tag{17}$$

Then one can use Eq. (15) in a mixed state as well. The advantage of it is that all the information about the mixed quantum state is rolled into the density matrix and is separated from the operator. The meaning of elements of the density matrix is as follows: $\rho_{uu}$ and $\rho_{dd}$ are probabilities of the spin pointing up and down the "z" axis, respectively; $\rho_{ud} = \rho_{du}^*$ is related to the projection of the spin on the "x" and "y" axes. Below are a few special cases of the density matrix corresponding to: spin in the positive "x" direction, spin in the positive "z" direction, and unpolarized spins

$$\rho|_x = \begin{pmatrix} 1/2 & -1/2 \\ -1/2 & 1/2 \end{pmatrix}, \quad \rho|_z = \begin{pmatrix} 1 & 0 \\ 0 & 0 \end{pmatrix}, \quad \rho|_{unp} = \begin{pmatrix} 1/2 & 0 \\ 0 & 1/2 \end{pmatrix}. \tag{18}$$

The Hamiltonian $H$ is the operator corresponding to energy of the system. Heisenberg equation expresses [18] the evolution of any operator average value

$$\frac{d\langle O \rangle}{dt} = \frac{i}{\hbar} \langle [H, \hat{O}] \rangle, \quad (19)$$

where the commutation relation is defined as

$$[\hat{O}_1, \hat{O}_2] \equiv \hat{O}_1 \hat{O}_2 - \hat{O}_2 \hat{O}_1. \quad (20)$$

**States of carriers and secondary quantization**

Free carriers in a crystalline solid can be electrons or can be holes in filled bands [20]. They have states that are characterized by the quasi-momentum in the crystalline lattice $\mathbf{k}$, by the spin index $s$ and by the band index $b$. The spin indices span the values $u$ and $d$. The band index here can mean a conduction or a valence band in a semiconductor, or it can mean one of the subbands of a quantum confined structure (quantum well etc.). Thus a state of one free carrier in the crystal is denoted as $|\mathbf{k}, s, b\rangle$. These states are orthogonal and normalized.

The state with no particles (which are synonymous to free carriers here), "vacuum", is designated as $|0\rangle$. In order to describe the state of many particles which are indistinguishable fermions, one introduces the second quantized operators, also called creation and annihilation operators [19]. They are defined by the "ket" and "bra" vectors corresponding to these states:

$$|\mathbf{k}, s, b\rangle = a^+_{\mathbf{k},s,b} |0\rangle, \qquad \langle \mathbf{k}, s, b| = \langle 0| a_{\mathbf{k},s,b}. \quad (21)$$

Thus a state with one particle is obtained from vacuum by action of one creation operator, and action of the annihilation operator turns the state with one particle into vacuum.

In order to describe fermions, their anti-commutators

$$\{\hat{O}_1, \hat{O}_2\} \equiv \hat{O}_1\hat{O}_2 + \hat{O}_2\hat{O}_1 \tag{22}$$

must obey certain relations. In the present case the anti-commutation relations are determined by the orthonormality of the states:

$$\{a_{\mathbf{k}1,s1,b1}, a_{\mathbf{k}2,s2,b2}\} = 0, \qquad \{a^+_{\mathbf{k}1,s1,b1}, a^+_{\mathbf{k}2,s2,b2}\} = 0, \tag{23}$$

$$\{a_{\mathbf{k}1,s1,b1}, a^+_{\mathbf{k}2,s2,b2}\} = \delta_{\mathbf{k}1,\mathbf{k}2}\delta_{s1,s2}\delta_{b1,b2}. \tag{24}$$

Now a pure state of N particles can be obtained by N-times action of various creation operators:

$$|\Psi(N)\rangle = \prod_{i=1}^{N} a^+_{\mathbf{k}i,si,bi} |0\rangle. \tag{25}$$

The complete basis of a many-particle problem consists of all possible states like this with all possible numbers of particles.

Any operator represented in terms of single-particle states now can be re-cast for the many-particle problem by the following recipe: one has to substitute any "ket" vector by the corresponding creation operator, and any "bra" vector by the annihilation operator.

**Magnetic field**

We first consider interaction of the spin of a particle (electron or hole in a crystal) with the external magnetic field $\mathbf{B}$. The Hamiltonian of this interaction

$$H_{mag} = -\boldsymbol{\mu} \cdot \mathbf{B} \tag{26}$$

is proportional to the magnetic moment of the particle. That, in turn, is proportional to the spin of the particle

$$\boldsymbol{\mu} = G\mathbf{S} \equiv \frac{gq}{2m_e}\mathbf{S} = \frac{gq\hbar}{4m_e}\boldsymbol{\sigma} = \frac{g \cdot \text{sign}(q)\mu_B}{2}\boldsymbol{\sigma}, \tag{27}$$

where we used Eq. (12). Here the charge $q$ equals $e$ for holes and $-e$ for electrons. Mass $m_e$ is the mass of a free electron. The Lande factor $g \approx 2$ for free electrons and can be different and even negative for particles in a crystal. We have defined the gyromagnetic ratio $G$ by Eq. (27). Conventionally the constant, Bohr magneton, is introduced by

$$\mu_B = \frac{e\hbar}{2m_e}. \tag{28}$$

Thus the Hamiltonian is

$$H_{mag} = -G\frac{\hbar}{2}\boldsymbol{\sigma}\cdot\mathbf{B}, \tag{29}$$

This Hamiltonian is acting only on the spin states of the particles and therefore we suppress the band and momentum indices in the rest of the subsection. Summation over all particles with various bands with various values of the momentum is implied. From (12) we infer that, in the second quantization notation, the operators of spin projections are

$$\sigma_x = a_u^+ a_d + a_d^+ a_u, \tag{30}$$

$$\sigma_y = -ia_u^+ a_d + ia_d^+ a_u, \tag{31}$$

$$\sigma_z = a_u^+ a_u - a_d^+ a_d. \tag{32}$$

Therefore we will calculate the evolution of the population-polarization matrix of average values in the many-particle quantum state

$$p_{s1s2} = \langle a_{s1}^+ a_{s2} \rangle, \tag{33}$$

which corresponds to the Hermitean conjugate of the density matrix, $\rho^*$ of a single-particle quantum state. The elements of this matrix are the quantities of practical interest that are measurable and pertinent to spin dynamics. The average values of the total spin for the ensemble of particles is given by

$$\langle \sigma_x \rangle = p_{ud} + p_{du}, \tag{34}$$

$$\langle \sigma_y \rangle = -ip_{ud} + ip_{du}, \tag{35}$$

$$\langle \sigma_z \rangle = p_{uu} - p_{dd}, \tag{36}$$

The total magnetization of particles is obtained from them via Eq. (27). The total number of particles is

$$N = p_{uu} + p_{dd}. \tag{37}$$

Now we can also express the energy of an ensemble of spins in the magnetic field. For example, for electrons in the magnetic field only along the "z" axis,

$$\langle H \rangle = \frac{g\mu_B}{2}(p_{uu} - p_{dd})B_z. \tag{38}$$

This means that if the field is along the positive "z" direction ("up"), then the lowest energy is achieved when the spin points "down", so that the magnetic moment is "up" and is aligned with the magnetic field.

The Heisenberg equation (19) with the magnetic field Hamiltonian yields

$$\frac{dp_{uu}}{dt} = i\frac{G}{2}(B_x - iB_y)p_{ud} - i\frac{G}{2}(B_x + iB_y)p_{du}, \tag{39}$$

$$\frac{dp_{ud}}{dt} = i\frac{G}{2}(B_x + iB_y)(p_{uu} - p_{dd}) - iGB_z p_{ud}, \tag{40}$$

$$\frac{dp_{du}}{dt} = \frac{dp_{ud}^*}{dt} \tag{41}$$

$$\frac{dp_{dd}}{dt} = -\frac{dp_{uu}}{dt}. \tag{42}$$

These are the Bloch equations for the non-interacting spins. By using the expressions for the projections of the spin on the axes (34)-(36), the Bloch equations (39)-(42) can be rendered in a form

$$\frac{d\langle\sigma_x\rangle}{dt} = G\langle\sigma_y\rangle B_z - G\langle\sigma_z\rangle B_y, \tag{43}$$

$$\frac{d\langle\sigma_y\rangle}{dt} = G\langle\sigma_z\rangle B_x - G\langle\sigma_x\rangle B_z, \tag{44}$$

$$\frac{d\langle\sigma_z\rangle}{dt} = G\langle\sigma_x\rangle B_y - G\langle\sigma_y\rangle B_x, \tag{45}$$

or in a real space vector form (42)-(45) turn to

$$\frac{d\langle\boldsymbol{\sigma}\rangle}{dt} = G\langle\boldsymbol{\sigma}\rangle \times \mathbf{B} \tag{46}$$

which is the familiar form of the Bloch equations [20] with its isotropic nature explicitly manifested. These equations describe precession of the magnetic moment of particles around the direction of the magnetic field.

**Kinetic energy**

The Hamiltonian of the kinetic energy has a very simple form – diagonal in the basis of the momentum states and independent of the spin

$$H_{kin} = \sum_{\mathbf{k},s,b} E_b(k) a^+_{k,s,b} a_{k,s,b}. \tag{47}$$

The energy of a particle in the band sometimes can be approximated by a parabolic dependence with a certain mass in a band

$$E_b(k) = \frac{\hbar^2 k^2}{2m_b}. \tag{48}$$

This mass in each band is in general different from the free-electron mass in vacuum.

We will be seeking equations of evolution for populations (probabilities of occupation) of states that are expressed by two-operator averages

$$p_{k,s1,s2,b1,b2} = \langle a^+_{k,s1,b1} a_{k,s2,b2} \rangle. \tag{49}$$

Evolution caused by this Hamiltonian is calculated to be

$$\frac{dp_{k,s1,s2,b1,b2}}{dt} = \frac{i}{\hbar}(E_{b1}(k) - E_{b2}(k)) p_{k,s1,s2,b1,b2}. \tag{50}$$

As we see, this Hamiltonian does not contribute to evolution of spin states, but rather only to inter-band terms. These terms are usually significant for optical properties of semiconductors. The spin state, on the other hand, can be detected by the manner it causes polarization dependence of optical interaction.

**Coulomb interaction**

The treatment of the Coulomb interaction between free carriers follows closely Ref. [14]. The energy of the electrostatic Coulomb interaction between two particles with charges $q_1$ and $q_2$ in 3 dimensions is

$$V(r) = \frac{q_1 q_2}{4\pi\varepsilon_0 \varepsilon_b r}, \tag{51}$$

where $\varepsilon_b$ is the background dielectric constant of the material, and $r$ is the distance between particles. The Hamiltonian operator for the ensemble of particle is obtained by adding up the interaction energy of all particles at various positions and in various bands and spin states:

$$H = \frac{1}{2} \sum_{s1,s2,b1,b2} \int d^3r_1 d^3r_2 \psi^*_{s1,b1}(r_1) \psi_{s1,b1}(r_1) V(|r_1 - r_2|) \psi^*_{s2,b2}(r_2) \psi_{s2,b2}(r_2). \tag{52}$$

The Fourier transform is required to express the Hamiltonian via momentum states.

$$f(r) = \sum_q f(q) e^{iq \cdot r} .\tag{53}$$

The sum over the momentum states means the integral over the $D$ non-confined dimensions, $L$ is the integration length in these dimensions.

$$\sum_q \equiv \int \frac{d^D q}{(2\pi)^D} L^D .\tag{54}$$

For example in a quantum well, it is a 2-dimensional integral, and the integration over the third dimension is replaced by the summation over subbands formed in the quantum well. At this point we also turn the wavefunctions into creation and annihilation operators.

$$H_{Cou} = \frac{1}{2} \sum_{k1,k2,q \neq 0, s1,s2,b1,b2} V(q) a^+_{k1+q,s1,b1} a^+_{k2-q,s2,b2} a_{k2,s2,b2} a_{k1,s1,b1} .\tag{55}$$

In the derivation one re-orders the operators to the normal ordering, i.e. all creation operators are to the left of all annihilation operators. This ordering means a choice of zero interaction energy in a vacuum state. Also the terms with $q = 0$ have been cancelled with the contribution of the crystal lattice: the uniformly spread positive ions counteract the repulsion of electrons.

The Fourier transform of the Coulomb energy (51) in 2 dimensions is

$$V(q) = \frac{q_1 q_2}{2 \varepsilon_0 \varepsilon_b L^2 q} .\tag{56}$$

The Coulomb interaction Hamiltonian contains 4-operator products, while we have seen before that 2-operator products correspond to quantities of practical significance. In the following we will be making the Hartree-Fock approximation: substituting the average values of the 4-operator product by the 2-operator product averages as follows

$$\langle a^+_1 a^+_2 a_3 a_4 \rangle = \langle a^+_2 a_3 \rangle \langle a^+_1 a_4 \rangle - \langle a^+_1 a_3 \rangle \langle a^+_2 a_4 \rangle + \delta \langle a^+_1 a^+_2 a_3 a_4 \rangle .\tag{57}$$

The first two terms in (57) are usually referred to as the direct and the exchange interaction. In many cases it is possible to also approximately account for the last term in (57) called "correlation". From all the 2-operator averages, we will be interested in ones with the same momentum. In other words, in this approximation we will neglect the correlations between states with different momenta which are responsible, for example, for screening. Note that this Hartree-Fock approximation in Semiconductor Bloch equations is different from the one made in the Feynman diagram technique: it corresponds to the infinite sum of all "ladder" diagrams in all orders of perturbation rather than just the first order diagrams.

The Coulomb interaction energy in the Hartree-Fock approximation is determined by the exchange interaction, and corresponds to attraction between electrons or between holes

$$E_{Cou} = -\sum_{k,q \neq 0, s1,s2,b1,b2} V(q) p_{k+q,s1,s2,b1,b2} p_{k,s2,s1,b2,b1} . \tag{58}$$

To understand the meaning of this expression, let us write out the spin states for one band

$$E_{Cou} = -\sum_{k,q \neq 0} V(q) \left[ p_{k+q,uu} p_{k,uu} + p_{k+q,ud} p_{k,du} + p_{k+q,du} p_{k,ud} + p_{k+q,dd} p_{k,dd} \right]. \tag{59}$$

From this expression we see that the state with all spins in the same direction, e.g. the only non-zero element $p_{k,uu} = 1$, has a lower energy than the state with unpolarized spins, the only non-zero elements $p_{k,uu} = p_{k,dd} = 0.5$. It proves that the Coulomb interaction favors the ferromagnetic state of carriers.

**Semiconductor Bloch equations**

Traditionally in literature (see e.g. [14]) semiconductor Bloch equations are used to describe interaction with electromagnetic field causing interband transitions. Here we apply them to the spin dynamics (where the classical Bloch equations originated). The new element here is the account of exchange Coulomb interaction which is responsible for ferromagnetism.

The Hamiltonian of Coulomb interaction (55) gives rise to additional contributions to the dynamics of spins. Using the Heisenberg equation (19), we express the derivatives of the 2-operator averages via 4-operator averages. The latter ones are approximated by Hartree-Fock as products of 2-operator averages as in (57). In the end, we obtain for the Coulomb part of the evolution

$$\frac{dp_{k,s1,s2,b1,b2}}{dt} = \sum_{q \neq 0, s, b} \frac{iV(q)}{\hbar} \left[ p_{k,s1,s,b1,b} p_{k+q,s,s2,b,b2} - p_{k+q,s1,s,b1,b} p_{k,s,s2,b,b2} \right]. \qquad (60)$$

The terms beyond the Hartree-Fock approximation are responsible for screening of the Coulomb potential and for carrier-carrier collisions. We will re-insert these terms in a semi-phenomenological manner later. The discarded contributions are also responsible for the higher-order correlations between particle states. Accounting of these terms via higher orders of 4, 6, etc. operator products makes the description much more complicated. At the same time it is not clear that it leads to a better agreement with the experiments.

Note that the Hartree-Fock terms of the Coulomb interaction in (60) cancel in one special case, when the spin state is independent of the momentum state for each band. Then $p_{k,s1,s2,b1,b2}$ can be factorized into the internal and the momentum parts. Therefore we can expect to see a non-trivial evolution due to Coulomb interaction only when the spin state is different for different momenta.

Thus Coulomb interaction causes the equations for populations and polarizations to be a large coupled set of nonlinear ordinary differential equations. To gain insight into the meaning of these equations we again write them for one band and explicit spin notation and re-insert the magnetic field terms:

$$\frac{dp_{k,uu}}{dt} = \sum_{q \neq 0} \frac{iV(q)}{\hbar} \left[ p_{k,ud} p_{k+q,du} - p_{k+q,ud} p_{k,du} \right]$$
$$+ i\frac{G}{2}(B_x - iB_y) p_{k,ud} - i\frac{G}{2}(B_x + iB_y) p_{k,du}$$
(61)

$$\frac{dp_{k,ud}}{dt} = \sum_{q \neq 0} \frac{iV(q)}{\hbar} \left[ p_{k,uu} p_{k+q,ud} + p_{k,ud} p_{k+q,dd} - p_{k+q,uu} p_{k,ud} - p_{k+q,ud} p_{k,dd} \right]$$
$$+ i\frac{G}{2}(B_x + iB_y)(p_{k,uu} - p_{k,dd}) - iGB_z p_{k,ud}$$
(62)

$$\frac{dp_{k,du}}{dt} = \frac{dp_{k,ud}^*}{dt},$$
(63)

$$\frac{dp_{k,dd}}{dt} = -\frac{dp_{k,uu}}{dt}.$$
(64)

This is the final form of the Semiconductor Bloch equations that is used for simulations in this article. We see that the structures of the Coulomb terms and the magnetic field terms are somewhat similar. In fact, one can easily see that the following substitution recasts (61)-(64) into the form of traditional Bloch equations

$$G\mathbf{B}_{mod}(k) = G\mathbf{B} + \sum_{q \neq 0} \frac{V(q)}{\hbar} \boldsymbol{\sigma}_{k+q}.$$
(65)

The meaning of this is that the magnetic moments tend to align to enhance the external magnetic field, and henceforth the evolution is governed by the modified magnetic field $\mathbf{B}_{mod}$. For holes, the spins are in the direction of the magnetic field. For electrons, the spins are opposite to the magnetic field, since the charge and the gyromagnetic ratio $G$ are negative. Note that it is not a simple reduction to traditional Bloch equations, since the Coulomb enhancement term is dependent on the momentum.

**Fermi distribution and collision relaxation**

In thermal equilibrium the populations of fermions are given by the Fermi distribution

$$f(k,s,b) = \frac{1}{1+\exp\left(\frac{E(k,s,b)-\mu_b}{k_B T}\right)}, \tag{66}$$

where $E(k,s,b)$ is the energy of a quantum state including both magnetic field and Coulomb interaction energy, $\mu_b$ is the chemical potential which may be different in each band, $T$ is the temperature, and $k_B$ is the Boltzmann's constant. The chemical potential is calculated from the condition that the density in a band is

$$N_b = \frac{1}{L^D} \sum_{k,s} f(k,s,b). \tag{67}$$

Even though our system can be strongly non-equilibrium, the Fermi distribution is crucial in several instances, especially in the description of collisions.

Collisions as well as other non-Hamiltonian evolution processes (which we call "relaxation") are crucial for generation of a ferromagnetic state. Purely Hamiltonian evolution cannot change the average energy, and therefore cannot transfer the system from its initial state to the lowest-energy, ferromagnetic state. Only collisions let the ensemble of particles relax to the ferromagnetic state.

A very fruitful approximation to describe collisions is a rate-equation approximation for the change of the population [14]

$$\frac{dp_{k,s,s,b,b}}{dt} = -\gamma_b \left( p_{k,s,s,b,b} - f(E(k,s,b)) \right), \tag{68}$$

where $\gamma_b$ is the rate of collisions corresponding to a certain band $b$. Care must be taken to correctly describe relaxation of off-diagonal 2-product averages. The form of terms responsible for the removal of particles which is invariant relative to spin rotations is

$$\left.\frac{dp_{k,s1,s2,b1,b2}}{dt}\right|_{out} = -\frac{\gamma_{b1}}{2} p_{k,s1,s2,b1,b2} \qquad (69)$$

with all possible combinations of spin and band indices. The terms corresponding to arrival of particles in the final state remain the same as in (68)

$$\left.\frac{dp_{k,s,s,b,b}}{dt}\right|_{in} = \gamma_b f(E(k,s,b)) \qquad (70)$$

with no contribution to off-diagonal elements.

**Screening**

As we mentioned before, we need to re-insert the screening effects into the equations. Following the method of [14] we modify the Coulomb potential to the screened Coulomb potential

$$V_s(q) = \frac{V(q)}{\varepsilon(q)}. \qquad (71)$$

In the "plasmon-pole approximation" it yields [14]

$$V(q) = \frac{e^2}{2\varepsilon_0 \varepsilon_b L^2} \cdot \frac{1 + q^2 Sc}{\kappa + q + q^3 Sc}, \qquad (72)$$

where the inverse Debye plasma screening length

$$\kappa = \frac{e^2 \sum_b m_b f_b(k=0)}{2\pi \varepsilon_0 \varepsilon_b \hbar^2}, \qquad (73)$$

and the screening constant arising for the electron-hole plasma

$$Sc = \frac{C\sum_b m_b f_b(k=0)}{16\pi \sum_b m_b N_b}, \qquad (74)$$

and C is taken between 1 and 4 for various materials.

**Injection and removal by electric current**

Depending on the nature of a problem at hand one might consider either a closed or an open ensemble. In the case of a closed ensemble no currents cross the boundaries of the system; therefore the number of particles or density is fixed. For the time-dependent problem, they are determined by the initial state. For the steady-state problem, one replaces one of the Semiconductor Bloch equations (which is a consequence of all other equations) by a condition of fixed density.

In the case of an open ensemble, particles are injected or removed from the system; and the equations are to be modified accordingly. Such a current will be treated as interaction with a large reservoir with temperature $T_R$ and chemical potential $\mu_R$ (determined by the density of particles in the reservoir). Energy of states in this reservoir may depend on all quantum variables $\tilde{E}(k,s_1,s_2,b)$ and, in fact, may represent a spin-polarized quasi-equilibrium state. The rate of injection from the reservoir in band $b$ is designated $r_b$. The removal of particles from the system for all bands is given by

$$\left.\frac{dp_{k,s1,s2,b1,b2}}{dt}\right|_{out} = -\frac{r_{b1}}{2} p_{k,s1,s2,b1,b2}\left(1 - f\left(\tilde{E}(k,s1,s2,b1)\right)\right). \qquad (75)$$

with all possible combinations of spin and band indices. The injection from the reservoir for each band is given by

$$\left.\frac{dp_{k,s1,s2,b,b}}{dt}\right|_{in} = r_b f(\tilde{E}(k,s_1,s_2,b))(1 - p_{k,s1,s2,b,b}). \qquad (76)$$

Note the last terms in parenthesis, which correspond to current blocking due to the fermionic nature of the particles. The total current per unit area is calculated as a sum of current contributions over all quantum states

$$I = e \sum_{k,s,b} \left( \frac{dp_{k,s,s,b,b}}{dt}\bigg|_{in} + \frac{dp_{k,s,s,b,b}}{dt}\bigg|_{out} \right), \quad (77)$$

while the spin component of the current as

$$I_j = e \sum_{k,s1,s2,b} (\sigma_j)_{s1,s2} \left( \frac{dp_{k,s,s,b,b}}{dt}\bigg|_{in} + \frac{dp_{k,s,s,b,b}}{dt}\bigg|_{out} \right), \quad (78)$$

where the index j can mean x, y, or z and $(\sigma_j)_{s1,s2}$ are the elements of the corresponding Pauli matrices. We define the spin gain of the transistor for each of the spin projections as the ratio of the integrated collector and base currents over the clock cycle

$$Gain_j = \frac{\int I_{j,collect}(t)dt}{\int I_{j,base}(t)dt}. \quad (79)$$

## IV Numerical results

In this section we apply the above mathematical model to simulate spin dynamics in a semiconductor and, in particular, operation of a spin gain transistor. The material system we approximate here is a GaAs quantum well with AlGaAs barriers. Though high Curie temperature was discovered in MnGaAs semiconductors, in this article we focus on a model of ferromagnetism arising from interaction of free carriers only. The model with spins of electrons localized in Mn ions is to follow. Since the operation of the device proposed here involves one type of carriers, we consider only one valence band here with effective mass $m_b = 0.5 m_e$ taken equal to the mass of heavy holes in GaAs. The material parameters used are: dielectric constant $\varepsilon_b = 12.9$, aggregated hole-hole and hole-phonon collision rate $\gamma = 3.5 \cdot 10^{12} s^{-1}$. The Lande factor for heavy holes is taken to be $g \approx -4$, which varies with the width of the quantum well, according to Ref. [21].

**Precession in the magnetic field**

To illustrate the workings of our numerical model we start with the precession of spins in the magnetic field of 5T in the negative "z" direction. The initial state is chosen to be Fermi distributed over the momenta with density $N = 5 \cdot 10^{11} cm^{-2}$ and with defined values of spin projections $\sigma_x = 1, \sigma_y = -2, \sigma_z = 3 \times 10^{11}$. First we demonstrate the influence of the magnetic field only on non-interacting particles, Figure 2. We see a uniform circular precession of the spin around the vector of the magnetic field. Now we introduce the Coulomb interaction and collisions to the numerical model. If the temperature is high (e.g. T=89K, see Figure 3), all the projections of the spin are damped to a steady state value of zero for the x- and y-components, and a small value of the z-component proportional to the magnetic field. This is characteristic for a paramagnetic state above the Curie temperature. If we now set the temperature to a low value (T=8.9K, see Figure 4), the x- and y-components are still damped to zero, while the z-component grows to a large fraction of density. This is characteristic of a ferromagnetic state.

**Ferromagnetic conditions**

Our model also demonstrates spontaneous magnetization – the one without the external magnetic field. The evolution to a spontaneously magnetized state is presented in Figure 5. We pick as an initial condition a Fermi distribution of carriers with a slightly higher chemical potential for spins in "up" z-direction. The simulation shows an increase of magnetization to a finite steady state value.

Next we explore the dependence of the steady-state magnetization on temperature and density, Figure 6. At a relatively low density, $N = 2 \cdot 10^{11} cm^{-2}$, all the spins are aligned at very low temperatures. As the temperature grows, magnetization decreases, until it turns to zero for all temperatures higher than the Curie temperature, 15K in this case. The temperature dependence, Figure 6, of the magnetization is consistent with the known results for ferromagnets [20]. For a higher density $N = 10^{12} cm^{-2}$, the Curie temperature becomes 27K. That means that at a fixed temperature, a ferromagnetic transition can be caused by increase of density.

However the dependence is not monotonic, as wee see from the phase diagram of magnetization, Figure 7. The ratio of magnetization per unit area divided by the density of holes per area is plotted here. The ferromagnetic state is destroyed, if the density increases above approx. $N = 2.2 \cdot 10^{12} cm^{-2}$.

There are two factors that explain this behavior. First, screening becomes stronger as the density increases. Second, as the carrier gas becomes degenerate, lower energy states become filled for both values of spin. Then a smaller the fraction of carriers, those close to the Fermi level, has different occupation of spin states, and thus their contribution to the effective magnetic field (65) decreases.

Some insight into the results of the simulation can be gained via analytical estimates for the conditions of a ferromagnetic state. In the approximation of a constant Coulomb potential $V$ between all particles, the exchange interaction causes the splitting $\pm\Delta$ of the potential energy for the particles with spin up and spin down. It is, according to (65),

$$\Delta = \frac{V \langle \sigma_z \rangle}{2}, \tag{80}$$

where $\langle\sigma_z\rangle$ is the average spin projection per unit area. Let us define as a border of Fermi degeneracy the density at which the chemical potential is at the band edge. For two dimensions it results in

$$N_{\text{deg}} = \frac{mk_BT\ln(2)}{2\pi\hbar^2}.\qquad(81)$$

For a degenerate Fermi gas, the condition of ferromagnetism is given by the Stoner criterion, in which DOS is the density of states for both spin states at the Fermi energy:

$$V > \frac{1}{\text{DOS}} = \frac{\pi\hbar^2}{m}.\qquad(82)$$

This criterion defines whether a ferromagnetic state exists at low temperatures.

For the limit of non-degenerate Fermi gas one can obtain the relation between the splitting and the difference of the populations $\langle\sigma_z\rangle$ with spin up and spin down

$$\langle\sigma_z\rangle = \frac{VN}{2}\tanh\left(\frac{\Delta}{k_BT}\right).\qquad(83)$$

Eqs. (80) and (83) have a non-trivial solution for splitting if

$$N > \frac{2k_BT}{V}.\qquad(84)$$

It defines the upper temperature bound for magnetization in Figure 7.

Simulations also show that the Curie temperature is higher for materials with smaller $\varepsilon_b$, in agreement with (84). Thus one can expect a high Curie temperature in, for example, GaNa, where the dielectric constant is lower. The reason for this is a stronger Coulomb exchange interaction which promotes ferromagnetism.

Eq. (82) shows that a higher mass of the carriers is favorable for ferromagnetism, again in agreement with our simulations. This is consistent with the fact that a higher Curie temperature has been observed in materials with hole-mediated ferromagnetism that with electron-mediated ferromagnetism, since electrons typically have a lower mass. As we stressed before, the Curie temperatures in the present model are lower than observed in MnGaAs, because the present model does not include localized spins of magnetic impurities.

**Spin gain transistor**

Then we simulate the dynamics of a spin gain transistor which was qualitatively described in Section II. In the beginning of the clock cycle, 0ps, one has a low density $N = 2 \cdot 10^{11} cm^{-2}$ of holes in the base, see Figure 8. After 0ps, a voltage applied to the emitter causes the emitter-base current; the base density increases. Since this current is not magnetized, see Figure 9, the magnetization in the base is zero. However the density is increased above the ferromagnetic transition. Between 5ps and 7ps a pulse of a base current is applied which is ~0.25 polarized in the z-direction. It does not produce appreciable density change but triggers spontaneous magnetization from 7ps to 14ps. Finally the voltage is applied to the collector to extract the polarized carriers from the base. From 18ps to 25ps, the density in the base as well as the fraction of the magnetization decreases. The base-collector current is almost entirely spin-polarized. Spin current gain of ~1100 is calculated for this case. The practical limit to gain is set by the spin fluctuations; their consideration is beyond the scope of this paper.

## V Summary

In this article we have examined operation of a spin gain transistor. Its principle of operation is based on the carrier control of ferromagnetism and control of spontaneous magnetization by a weak current. It enables obtaining a strong spin-polarized current which is a replica of the control current with a gain in spin component of the current of more than 1000. We developed a theory and a numerical simulation scheme to describe such a transistor based on Semiconductor Bloch Equations. We describe within the same formalism precession of spins in a magnetic field, ferromagnetic ordering without the magnetic field, temporal evolution of charges and spins in a spin transistor. Results of simulations for a generic material system are presented which validate that the mathematical model describes the qualitative trends of ferromagnetism in semiconductors and demonstrates the principle of operation of the spin gain transistor.

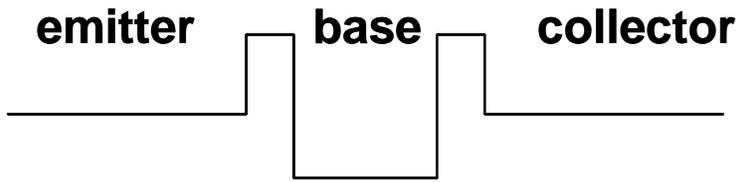
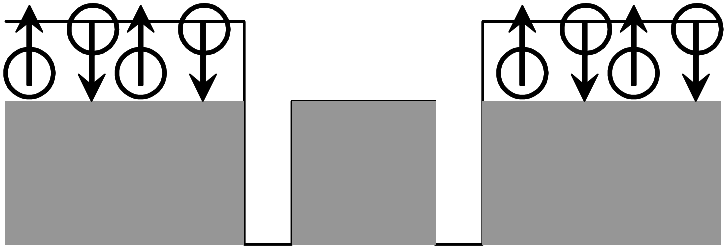
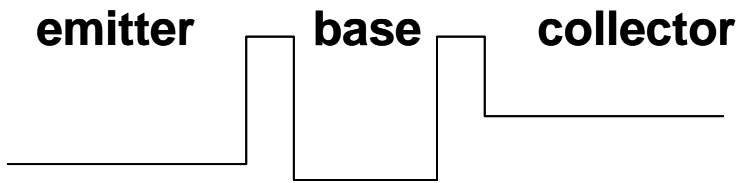
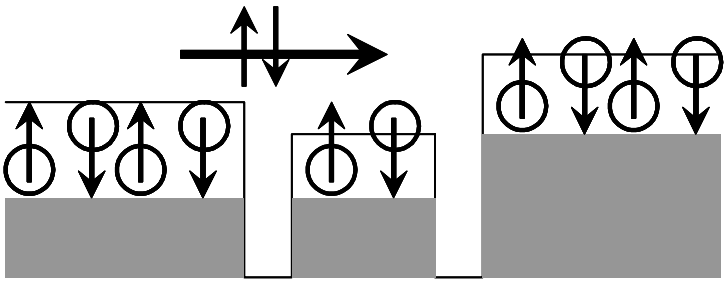

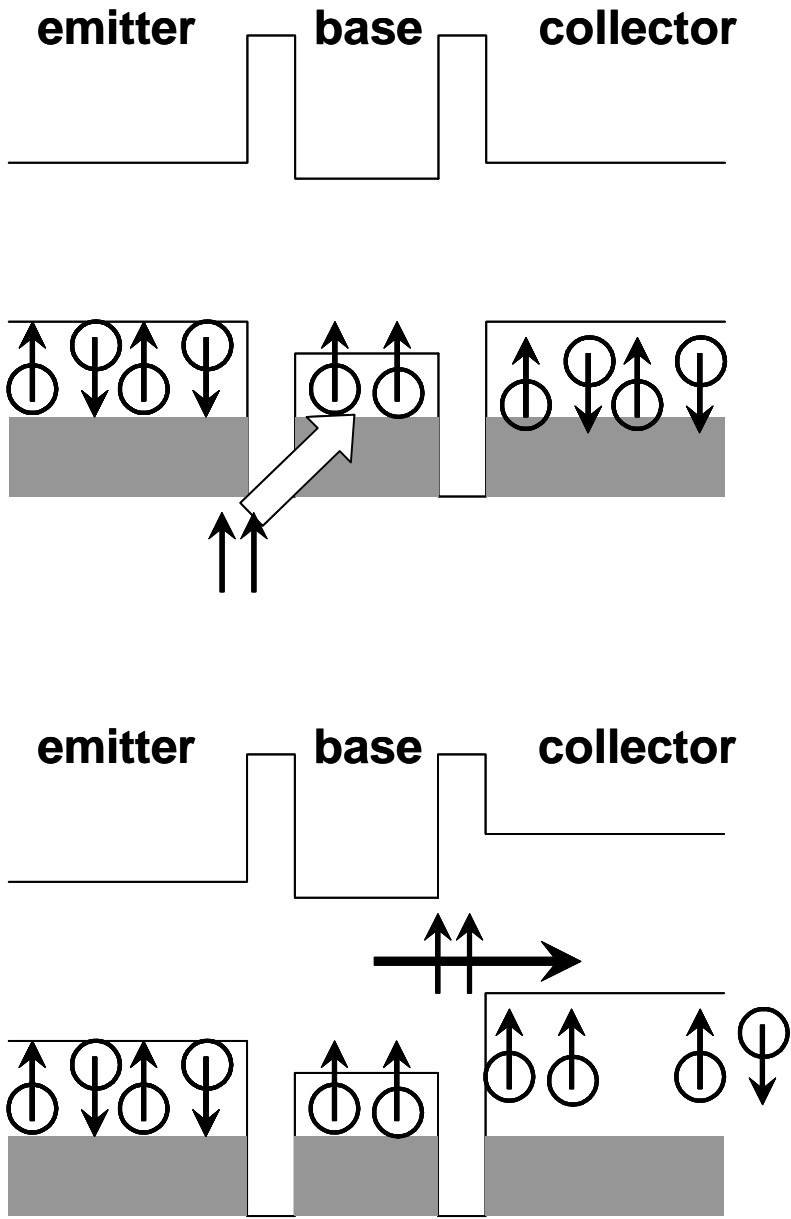

Figure 1. Band diagrams of a spin gain transistor and stages of its operation. Holes and their spin direction are schematically shown. a) Initial, no magnetization; b) current is injected from the collector to the base, ferromagnetic condition; c) spin-polarized base current injected, ferromagnetic state; d) output spin-polarized current.

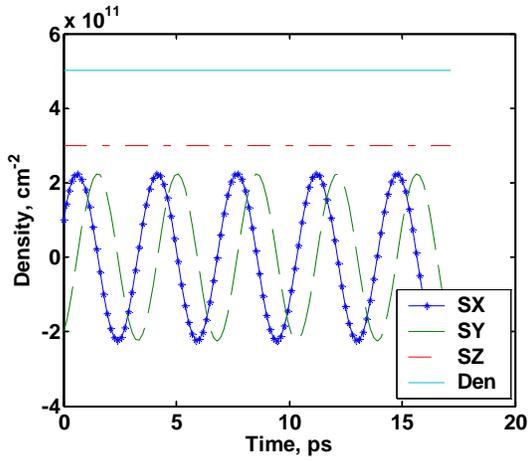

**Figure 2. Precession of spins in the magnetic field in a hypothetical case without relaxation, with magnetic field Bz=-5T. Here and after "Den" is the density of holes, "SX", "SY" and "SZ" are sums of the spin projections (per unit area).**

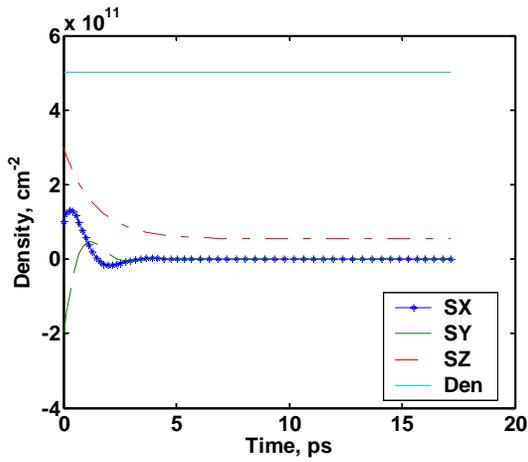

**Figure 3. Damped precession in the magnetic field Bz=-5T, T=89K.**

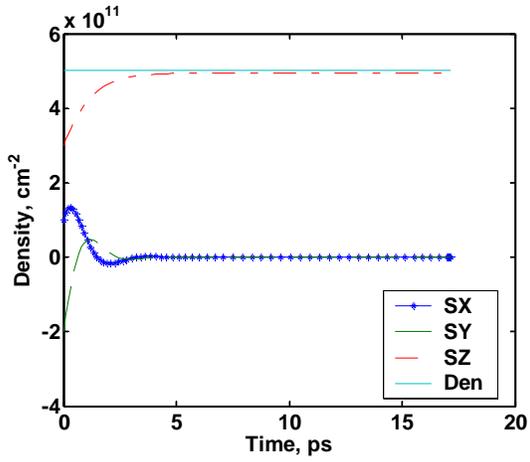

**Figure 4. Generation of a ferromagnetic state in magnetic field Bz=-5T, T=8.9K.**

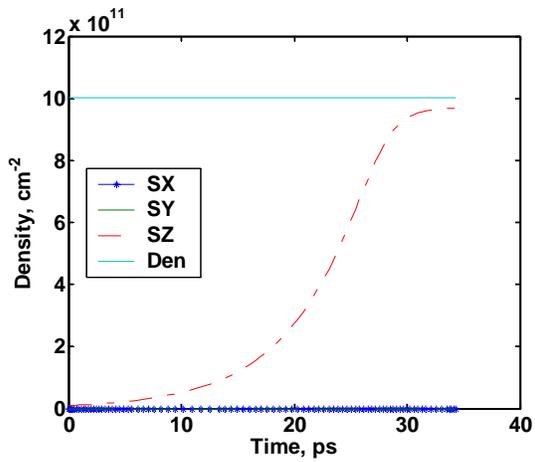

**Figure 5. Generation of spontaneous magnetization with B=0T starting with a small perturbation of an unpolarized spin distribution.**

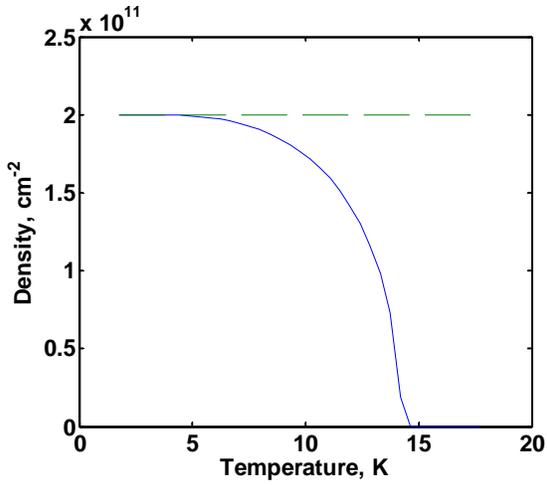

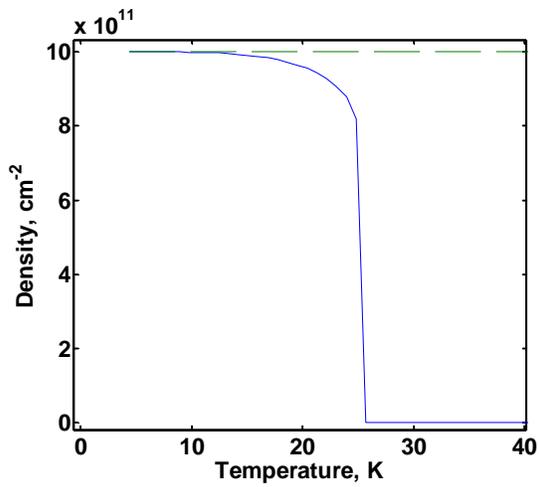

**Figure 6. Dependence of spontaneous magnetization on temperature. Solid line – magnetization, dashed line - density. Carrier density a) $0.2*10^{12}$cm$^{-2}$; b) $10^{12}$cm$^{-2}$.**

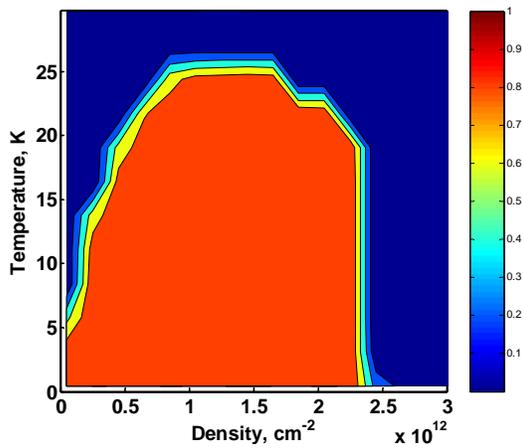

**Figure 7. Phase diagram – dependence of spontaneous magnetization on density and temperature.**

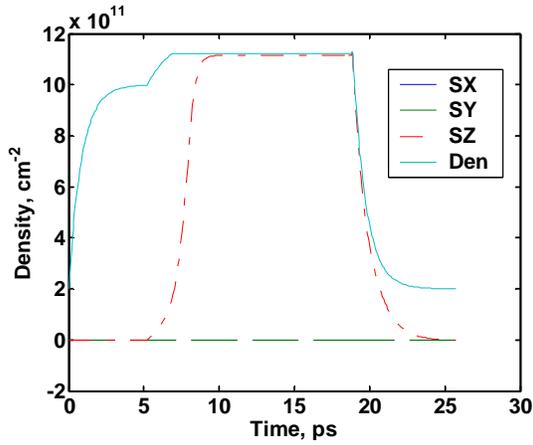

**Figure 8. Evolution of density and magnetization over a clock cycle in a spin gain transistor, T=8.9K.**

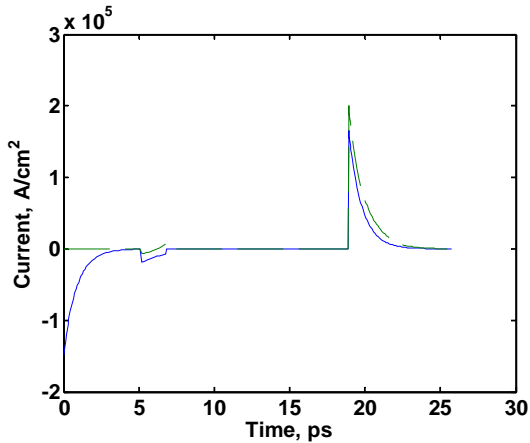

**Figure 9. Evolution of the total current (solid line) and the spin-polarized component of the current (dashed line) over a clock cycle in a spin gain transistor. Parameters same as Figure 8.**